\documentstyle{article}
\begin{document}

\newcommand{\be}{\begin{equation}}
\newcommand{\ee}{\end{equation}}

\centerline{\bf ENERGY DOMINANCE  AND THE}
\centerline{\bf HAWKING ELLIS VACUUM CONSERVATION THEOREM} 

\vskip 1 cm

\centerline{\bf Brandon Carter}

\vskip 0.6 cm
\centerline{LUTH, Observatoire de Paris}
\centerline{{\bf 92} 190 Meudon, France.}
\vskip 0.6 cm

\parindent= 0 cm

{\it A contribution to Stephen Hawking's 60th birthday
workshop on the Future of Theoretical Physics and Cosmology,
Cambridge, January 2002}

\hskip 1 cm

\parindent= 0 cm

{\bf Abstract.}  At a time when uninhibited speculation about
negative tension -- and by implication negative mass density --
world branes has become commonplace, it seems worthwhile
to call attention to the risk involved in sacrificing traditional
energy positivity postulates such as are required for the
classical vacuum stability theorem of Hawking and Ellis.
As well as recapitulating the technical content of this
reassuring (when applicable) theorem, the present article provides 
a new, rather more economical proof.

\bigskip
\parindent= 0 cm
{\bf 1.  Introduction}.
\medskip \parindent= 1 cm

Although overshadowed by other more recent contributions --
such as the no-boundary recipe for creation of an entire universe -- 
one of the most obvious subjects for reminiscence on the auspicious 
occasion of this 60th birthday celebration for Stephen Hawking is his 
central role in the foundation of classical black hole theory as a 
mathematical discipline in the late 1960's and early 1970's. It is 
remarkably fortunate that it has been possible, thirty years later, to 
assemble so many of the other protagonists in that memorable collective 
enterprise, including Roger Penrose, Werner Israel, Jim Hartle, Kip Thorne, 
Jim Bardeen, Charles Misner, Martin Rees, Gary Gibbons, and particularly 
George Ellis, the coauthor with Stephen of their landmark treatise 
``The Large Scale Structure of Space Time'' (Cambridge 1973)~\cite{HE73}, 
which remains unsuperseded as the definitive reference on this subject, 
having been published just at the time when, still under Stephen's 
leadership, the emergence of black hole thermodynamics diverted the 
main thrust of progress in black hole theory from classical to quantum
aspects. 

A central problem in the classical theory to which Stephen made a
particularly masterly and important contribution was the question of black 
hole equilibrium states, as described in my recent introductory historical
overview~\cite{C99}. Due to (local rather than global) limitations of both 
space and time I shall not attempt to take up the challenge of providing 
for this occasion a more general historical review of (dynamical as well as 
stationary) classical black hole theory and of Stephen's leading role therein. 
Before coming to the main point of this brief contribution I would just like 
to advertise the  previously published personal reminiscences of some of 
those involved~\cite{I87,T94} and to emphasize that a serious student of the 
subject could still not do better than to start by working through the 
relevant sections of Hawking and Ellis (1973)~\cite{HE73}, as reprinted in 
its original form, which was crafted so well that, even after all these 
years, no new edition or replacement has been required. (The only small 
point on which I am conscious of the need for any caveat concerns the 
questionable assumptions in a heuristic energy extraction argument 
invoked~\cite{HE73} -- on page 328 -- as a step towards the important 
conclusion that  a stationary non-rotating black hole configuration should 
be strictly static. As briefly described in the cited overview~\cite{C99}, 
a mathematically sound basis for this conclusion has finally been 
provided~\cite{SW91,CW93,SW93} by much more recent work under the 
leadership of Bob Wald.)

What I would like to do instead on this occasion is to draw attention
to another quite distinct area (not specifically concerned with black
holes) to which the treatise of Hawking and Ellis (1973)~\cite{HE73}
made a particularly significant contribution, namely the question of 
energy positivity conditions and their use in establishing the stability 
of the vacuum against spontaneous creation processes. This question has 
recently become rather topical in view of the current 
fashionability~\cite{GRS00} of higher dimensional theories involving
what are euphemistically described as ``negative tension branes''.

Generalising the familiar case of an ordinary membrane with a support 
surface having two space dimensions (as well as one dimension of time), the 
term ``brane'' has come to be used for the limit of a system confined 
within a neighbourhood of relatively small thickness about a supporting 
worldsheet surface of arbitrary dimension, as exemplified by  the special 
cases of a ``string'' with only a single space dimension.  Everyday 
architectural experience shows that negative tension can perfectly well be 
sustained by a supporting column of sufficient thickness relative to its 
length, but it can be shown quite generally~\cite{C95} that in the small 
thickness limit to which the term ``brane'' refers, negative
tension will always be accompanied by instability against lateral --
i.e. ``wiggle'' type -- perturbations, an effect that can be dramatically
demonstrated (as I found quite inadvertently!) by pushing too hard on an
ordinary thin pointing rod (such as the one rashly provided to me by
Gary Gibbons on this occasion).

However it is not this ordinary destabilising kind of negative tension that 
is involved in the higher dimensional brane world models of the kind 
considered by authors such as Gregory, Rubakov and Sibiryakov~\cite{GRS00}, 
but something far more exotic and dangerous. There are two reasons why the 
trouble with these models does not arise from ordinary wiggle instabilities. 
The first is that although a few authors have actually considered braneworlds 
having the full range~\cite{BCMU01} of lateral degrees of freedom as in 
ordinary branes, the majority, starting with Randal and Sundrum~\cite{RS99}
have preferred to postulate that the relevant supporting worldsheets should  
be of bounding orbifold type, or subject to a reflection symmetry, which 
suppresses all the lateral degrees of freedom, thus making it rather 
misleading to use the term ``brane'' at all in this context. The other reason 
is that, unlike the non relativistic membranes and strings (or rods) that are 
familiar in everyday life, whose tension (whatever its sign) has a magnitude 
that is small (in relativistic units) compared with their mass densities, 
the kinds of brane that occur in modern higher dimensional theories are 
typically of the Dirac type characterised by a tension that is approximately 
equal to the corresponding surface mass density. Thus, in such a brane,  
negativity of the tension has the alarming implication that the mass density 
itself should also be negative. This avoids the risk of lateral instability 
(which arises only when the signs are opposite) but at the expense of 
something that is far more frightenning, namely a flagrant violation of the 
principle that Hawking and Ellis (1973)~\cite{HE73} baptised as the ``weak 
energy condition'', and thus a fortiori of the usual ``dominant energy'' 
condition described below.

The conventional wisdom is that admissible theories must respect this kind of
energy positivity condition (at least on a classical macroscopically averaged
level) in order to avoid instability of the vacuum against a runaway process
of creation of positive and negative mass particles.  It is perhaps
conceivable that the situation might be saved by specific restrictions
forbidding the excitation of the degrees of freedom that would be involved in
such a runaway process, but in the recent words of Ed Witten~\cite{W00}, it 
seems more ``likely that physics with violation of the weak energy condition 
is unstable.''  To be more specific, it is clear that in the absence of such 
an energy condition it would no longer be possible to invoke what Hawking and
Ellis referred~\cite{HE73} to simply as the ``Conservation Theorem'', a 
result that might be described more specifically as the Vacuum Conservation 
Theorem, whose upshot (when applicable) is effectively that, at a classical 
level, the vacuum must be stable against spontaneous matter creation 
processes. 

As well as recapitulating the technical content of this noteworthy vacuum 
stability theorem, whose original proof sprawled over pages 92 to 94 of 
Hawking and Ellis (1973)~\cite{HE73}, the main objective of this brief
contribution is to offer a modified derivation that is rather more concise.

\bigskip\parindent=0 cm
{\bf 2  The energy dominance condition}
\medskip\parindent = 1 cm

The Hawking Ellis {\it vaccuum conservation theorem} -- to the effect that 
one cannnot create something from nothing -- applies  to cases describable in
terms of classical fields characterised by a stress momentum energy density
tensor subject to a postulate that Hawking and Ellis (1973)~\cite{HE73}
referred to as the ``dominant energy'' condition. Following the 
recapitulation of the contents of this postulate immediately below in 
the present section, the formal statement and the proof (in a new  
technically simpler version) of the theorem itself will be given in 
the final section of this article.

In a spacetime characterised by a time orientable pseudo Riemannian metric 
with components $g_{\mu\nu}$ and Lorentz type signature such that the 
condition for a vector $u^\mu$ to be a timelike unit vector is
\be u^\mu u_\mu =-1 \, ,\label{1}\ee
the meaning of the postulate that a (symmetric) stress momentum energy 
density tensor $T^{\mu\nu}$ satisfies the energy dominance condition in 
question is that for {\it any} future directed timelike unit vector
$u^\mu$ the corresponding energy flux vector 
\be {\cal E}^\mu=-T^{\mu\nu} u_\nu  \label{2} \ee
should be non-spacelike, i.e.
\be  {\cal E}^\mu {\cal E}_\mu \leq 0\, ,\label{3}\ee
with future time orientation,  i.e.
\be  {\cal E}^\mu u_\mu\leq 0\, . \label{4}\ee
This latter requirement (\ref{4}) is evidently equivalent
to the requirement that the corresponding energy density
scalar be non negative, i.e.
\be T^{\mu\nu}u_\mu u_\nu\geq 0\, .\label{5}\ee

It is to be observed that if $t_\mu$ is any vector that is also timelike,
\be t^\mu t_\mu<0\, ,\label{6}\ee
then the condition (\ref{3}) that the energy flux vector ${\cal E}^ \mu $ 
should be non-spacelike implies that its contraction with $t_\mu$ can 
vanish only if the energy flux vector itself vanishes, i.e.
\be {\cal E}^\mu t_\mu= 0\hskip 1 cm\Rightarrow \hskip 1 cm
{\cal E}^\mu=0\, ,\label{7}\ee
and that if $t^\mu$ is past directed, i.e.
\be t_\mu u^\mu > 0\, ,\label {8}\ee
the contraction will in any case have the non-negativity property
\be {\cal E}^\mu t_\mu\geq 0\, ,\label{9}\ee
which is equivalent to the condition  that 
\be T^{\mu\nu}u_\mu t_\nu\leq 0\, ,\label{10}\ee
for any pair of respectively future and past
directed timelike vectors $u^\mu$ and $t^\mu$.

A further almost equally obvious consequence of the energy dominance
condition is that ${\cal E}^\mu$ cannot vanish for any
particular unit vector $u^\mu$ unless it vanishes for all such
unit vectors, which will happen only in a vacuum where $T^{\mu\nu}$
vanishes altogether, i.e.
\be {\cal E}^\mu=0\hskip 1 cm \Rightarrow \hskip 1 cm
T^{\mu\nu}=0 \, .\label{11}\ee
It can thus be seen, by combining (\ref{7}) and (\ref{11}),
that according to the energy dominance condition the possibility for the 
contraction $T^{\mu\nu}u_\mu t_\nu$ to vanish for any timelike vectors 
$t^\mu$ and $u^\mu$ is excluded everywhere except in a vacuum, i.e.
\be T^{\mu\nu}u_\mu t_\nu=0\hskip 1 cm \Rightarrow \hskip 1 cm
T^{\mu\nu}=0 \, .\label{12}\ee

\bigskip\parindent= 0 cm
{\bf 3. The vacuum conservation theorem}
\medskip\parindent= 1 cm

Exploiting the existence (demonstrated in Section  6.4 of their
book~\cite{HE73}) of a globally well behaved time coordinate, $\tau$ -- 
i.e. a field with everywhere strictly timelike gradient $\tau_{;\mu}$ -- 
and assuming the validity of the ordinary local energy momentum 
conservation condition 
\be T^{\mu\nu}{_{{\bf ;}\nu}}=0\, ,\label{13} \ee
(using a semi colon for Riemannian covariant differentiation) Hawking and 
Ellis showed, in Section 4.3 of their book~\cite{HE73}, how the energy 
dominance postulate that has just been described can be used to derive a 
{\it  vacuum conservation theorem} whose purport is as follows: if the 
boundary of a compact causally well behaved space-time volume,  
${\cal V}$ say, consists just of an ``initial'' (but not necessssarily 
spacelike) vacuum hypersurface $\Sigma_{(0)}$ say -- i.e. a 
hypersurface  where $T^{\mu\nu}$ vanishes -- together with a future 
boundary hypersurface, $\Sigma_{(1)}$ say, that is {\it spacelike}, then 
the entire space time volume ${\cal V}$ will be characterised by the vacuum 
property, $T^{\mu\nu}=0$.

This result is obtainable as an immediate corollary of a lemma to the 
effect that the vacuum property will hold on the future boundary 
$\Sigma_{(1)}$ (and thus on the entire boundary): it evidently suffices to 
apply this lemma to the intersection of ${\cal V}$ with the past of a 
timelike hypersurface (given by a fixed global time $\tau$) through any 
point under consideration.

It is evident from (\ref{10}) and (\ref{12}) that to establish the required 
lemma, it will be sufficient to demonstrate the non-positivity, and hence 
the vanishing, of an integral of the (generically positive) form 
\be {\cal I}= \int_{\Sigma_{(1)}} {\cal E}^\mu t_\mu \, d\Sigma
\, ,\label{14}\ee
for some pair of repectively future and past directed timelike vector 
fields $u^\mu$ and
$t^\mu$ on the ``final'' hypersurface $\Sigma_{(1)}$ in question.

In order to do this, let us  consider the case for which $u^\mu$ is taken to 
be the unit future (i.e. outward) directed normal to the hypersurface. The 
corresponding normal surface element will then be expressible as 
$d\Sigma_\mu= -\ u_\mu\,d\Sigma$, so that we shall obtain
\be {\cal I}= \int_{\Sigma_{(1)}} T^{\mu\nu}t_\nu\, d\Sigma_\mu
\geq 0\, .\label{15}\ee

If $t^\mu$ is taken to be proportional to any one of the globally well 
behaved past directed timelike unit vector fields that can always be 
constucted (see Section 2.6 of Hawking and Ellis~\cite{HE73}) in any time 
orientable space time manifold, one can use Green's theorem to convert the 
surface integral (\ref{15}) to the form
\be {\cal I} = 
\int_{\cal V} \big(T^{\mu\nu} t_\nu\big)_{{\bf;}\mu}
\, d{\cal V}\, ,\label{17}\ee
as a consequence of the postulate that the vacuum condition should already 
be satisfied on the remaining ``initial'' part of the boundary of the 
relevant space time volume ${\cal V}$.

More specifically (relying on the the causal good behaviour postulate) let 
$\tau$  be the globally well defined time coordinate field invoked above.  
Then  -- as pointed out by Hawking and Ellis~\cite{HE73} in their Section 
4.3 -- the energy dominance condition ensures that in any compact space time 
region ${\cal V}$ there will be some finite positive constant, $C>0$, such 
that
\be | T^{\mu\nu} \tau_{{\bf;}\mu\nu}|\leq
CT^{\mu\nu} \tau_{{\bf;}\mu}\tau_{{\bf;}\nu} \, .\label{19}\ee
If we now choose the timelike vector in (\ref{17}) to be the gradient of a 
new exponentially related time coordinate $t$ according to a specification 
of the form
\be t_\mu=t_{{\bf ;}\mu}\, ,  \hskip 1 cm
C(t -t_{_\infty})=- {\rm e}^{-C\tau}\, ,\label{20}\ee
for some constant $t_{_\infty}$, 
then it can be seen from (\ref{13}) that the expression (\ref{17}) will 
reduce to the form
\be {\cal I} = 
\int_{\cal V} T^{\mu\nu} t_{{\bf;}\mu\nu}
\, d{\cal V}\, ,\label{22}\ee
with \be  t_{{\bf;}\mu\nu}={\rm e}^{C\tau}\big(\tau_{{\rm ;}\mu\nu}
-C\tau_{{;}\mu} \tau_{{;}\nu}\big)\, .\label{24}\ee
It then follows from (\ref{19}) that we shall have
\be T^{\mu\nu} t_{{\bf;}\mu\nu}\leq 0\, ,\label{25}\ee
and hence
\be {\cal I}\leq 0\, ,\label{26}\ee
which is compatible with the non-negative nature of the integral (\ref{15}) 
only if it vanishes.

This completes the proof of the lemma (and hence of the theorem) since, as 
noted above, the conclusion that the integral will vanish,
\be {\cal I}=0\, ,\label{27}\ee
implies {\it Q.E.D}, namely that, by  (\ref{10}) and (\ref{12}), the vacuum 
condition \be T^{\mu\nu} =0\, ,\label{28}\ee
will indeed have to be satisfied everywhere on the ``final'' hypersurface 
$\Sigma_{(1)}$ (and hence throughout ${\cal V}$).

\end{document}